\documentclass[intlimits,twoside,a4paper]{article}

\usepackage[cp1251]{inputenc}
\usepackage{multirow}
\usepackage{cmpj3}
\usepackage{bm}

\issue{2022}{25}{2}{23601}
\doinumber{10.5488/CMP.25.23601}

\title[Does the second critical-point of water really exist in nature?]%
{Does the second critical-point of water really exist in nature?%
}

\author[F. Hirata]{F. Hirata}

\address{Institute for Molecular Science, Okazaki, Aichi-444-8585, Japan 
}

\date{Received March 28, 2022, in final form May 18, 2022}
\begin{document}
	
\maketitle

\begin{abstract}

In the past decade, a literary phrase ``No man's land'' has been flooded in the scientific papers. The expression is used to describe a meta-stable region in the phase-diagram that cannot be accessed by experiments. It has been claimed based on the molecular dynamics (MD) simulation that there is a critical point, or the second critical point (SCP), in the ``no man's land,'' and it has created a big dispute in the field of science. It is proved in the present paper that the hypothesis of SCP is completely against the rigorous theorem of thermodynamics, referred as the \textit{Gibbs phase rule}. The reason why the simulations have found SCP erroneously is merely because
the method violates the requirement which all the statistical-mechanics treatments should satisfy
to reproduce the thermodynamics. That is the thermodynamic limit. It is clarified what is the identity of the ``liquid-liquid phase transition'' and SCP in pure liquids, \textit{discovered} by the simulations and by some experiments. In order to explain the physics of liquid-liquid phase transition observed experimentally in \textit{single }component liquids, a new concept is proposed.
\printkeywords

\end{abstract}

\section{Introduction}

The equilibrium and/or stability of fluid phases in the phase-diagrams, projected onto the ($P$, $T$) axes, is characterized by the sign of \textbf{$\left(\rd P/\rd V\right)_{T} $}: the stable region, $\left(\rd P/\rd V\right)_{T} \leqslant 0$; the unstable region, \linebreak$\left(\rd P/\rd V\right)_{T}>0$. The stable region is separated from the unstable region by a \textit{spinodal} \textit{line}. The stable region is further separated into stable and metastable regions by a \textit{coexistence} curve where the two fluid-phases coexist. The metastable region is located in the space between the two lines, i.e., coexisting and spinodal lines. The both curves merge into a point, referred to as the ``\textit{critical point},'' where $\left(\rd P/\rd V\right)_{T} =0$. Almost a parallel picture holds for a system consisting of more than one chemical components, each identified with the chemical potential, but the behavior of the phase diagram becomes more complicated~\cite{1}. The behavior of the phase diagram is subject to a theorem, referred to as the phase rule, proposed and proved strictly by J.~Willard Gibbs~\cite{2}. The phase rule governs the relation among the number of independent thermodynamic-variables and the number of phases realized. For the case of pure water, only three phases are allowed in nature, vapor, liquid, and ice, and a single critical-point appears between the two fluid phases. That was the common understanding among scientists before the paper by Pool et al. was published in Nature in 1992 \cite{3}. Based on the molecular dynamics (MD) simulation for a system consisting of a few hundred water-molecules, Pool et al. proposed a hypothesis named ``stability-limit conjecture'' that claims the existence of the second critical point (SCP) in the super-cooled region in the phase diagram of water \cite{3}. The experimental finding by Mishima et al., which shows phase separation in amorphous ice, was used as an empirical support of the conjecture \cite{4}. No wonder, the finding and conjecture have surprised many scientists, since the conjecture apparently conflicts with the Gibbs phase rule. So much effort by means of experiments as well as the molecular simulations was devoted enthusiastically to prove the conjecture \cite{5,6,7,8,9,10,11}. Nevertheless, the existence of SCP seems to be far from reality. In order to express such a situation, the community of SCP has used a phrase ``No mans land'' that refers to a meta-stable region in the phase diagram that cannot be accessed by experiments~\cite{7}. What is the nature that cannot be accessed by experiment, but just by MD? Why can they  say for sure that the SCP exists in nature without carrying out an experiment, which cannot be performed by a human being?

   In the present paper, it is proved based on the Gibbs phase rule that the SCP does not exist as an equilibrium or stable phase. It is argued that the appearance of the second critical point (SCP) is due to the artifact of the molecular simulation, that violates the thermodynamic limit as well as the ergodic theorem. It is pointed out that the ``phase separation'' claimed to be found experimentally for some liquid systems is nothing related to the SCP, but are the phenomena observed either for a liquid system consisting of two chemical species identified with different intramolecular chemical potentials, or for a molecular-scale process such as \textit{solvation} and \textit{structural fluctuation of protein}, rather than a phase behavior. (The meaning of the intramolecular chemical potential will be clarified later in section~\ref{sec4}).

\section{Proving non-existence of the second critical point based on the Gibbs phase rule}

The phase rule established by J. W. Gibbs is one of the most important theorems, that governs the relation between the number of equilibrium phases to which any thermodynamics system in nature may access and the number of chemical components included in the system~\cite{2}. Of course, the second critical point (SCP) by Pool et al. should be subject strictly to the rule, if such a point really exists in nature. The Gibbs phase rule can be written as,

\begin{equation}  r\leqslant n+2, 
\end{equation} 
where $r$ denotes the number of phases coexisting in the system, $n$ is the number of chemical components, each identified with the chemical potential~\cite{1}. According to the rule, a single component system may have three phases at most, because $n=1$ for a single component system. Thus, water in nature can have just \textit{three} phases at most, solid, liquid, and vapor. This was the common understanding until the second critical point (SCP) conjecture by Pool et al. was proposed. 

   Does the SCP conjecture reconcile with the Gibbs phase rule? The answer is ``No''. Suppose that such a point really exists. Then, four equilibrium fluid-phases, which can be distinguishable, should exist in the phase diagram: gas phase, the ordinary liquid phase in equilibrium with the gas phase, the two liquid phases separated by the spinodal line that creates the second critical point. Thus, the equilibrium thermodynamic-phases that may appear with changing the variables, $T$, $P$ and the density can be five all together, considering one phase for the solid state. On the other hand, the maximum number of variables allowed to a system consisting of a single chemical species, identified by the chemical potential, is \textit{three} according to the Gibbs phase rule. Thus, it is proved that the claim of the SCP violates the Gibbs phase rule. Now, it may be appropriate to review the Gibbs phase rule for some readers who are not familiar with the rule.  

  The proof of the phase rule is given repeatedly by many authors in standard textbooks of thermodynamics \cite{1,12,13}. Among the thermodynamic potentials, the Gibbs free energy ($G$) is the one that is most appropriate for the topics discussed here, since ordinary experiments to construct the phase diagram are made under the condition of pressure ($P$) and temperature ($T$) being constant, which of course are the canonical intensive variables of $G$. The Gibbs free energy is an extensive thermodynamic variable that satisfies the Euler theorem for a homogeneous function of the first degree, that is, 

\begin{equation}  G=\sum _{i}\mu _{i}  N_{i},
  \end{equation} 
where $N_i$ denotes the number of particles of a chemical species \textit{i }included in the system, and $\mu_{i} $ is the Gibbs free energy per particle of the chemical component $i$, called ``chemical potential''.  The chemical potential is also identified as a partial molar Gibbs free energy,

\begin{equation}  \mu _{i} =\left(\frac{\partial G}{\partial N_{i} } \right)_{T,P,j\ne i}.  \end{equation} 

   Let us suppose that the system is separated into two phases in contact, such as gas and liquid, or two liquids (solutions) phases with different concentrations. If one of the components, \textit{i,} is in equilibrium between the two phases specified by I and II, the equilibrium condition of the \textit{i}-th component between the two phases is expressed as $\mu _{i}^{I} =\mu _{i}^{II} $. If the chemical potential is different, i.e., $\mu _{i}^{I} >\mu _{i}^{II} $, the molecules in the system undergo diffusive motion in order to attain an equilibrium, and eventually satisfies the condition $\mu _{i}^{I} =\mu _{i}^{II} $ in equilibrium. Now, let us generalize the relation to a case in which a variety of phases including multiple chemical species are in contact. Let the number of chemical components that are identified with the chemical potential be $n$, and the number of different phases in contact be $r$. The entire system is equilibrated with the bath having constant temperature and constant pressure. The equilibrium condition can be expressed as

\begin{equation}  \begin{array}{l} {\mu _{1}^{I} =\mu _{1}^{II} =\cdots =\mu _{1}^{r} }, \\ {\mu _{2}^{I} =\mu _{2}^{II} =\cdots =\mu _{2}^{r} }, \\ {{\rm \; \; \; \; \; \; \; \; \; }\cdots \cdots }\,, \\ {\mu _{n}^{I} =\mu _{n}^{II} =\cdots =\mu _{n}^{r} }. \end{array}\label{eq4} \end{equation} 
Each chemical potential $\mu _{i}^{\Phi } $ ($\Phi =I,II,\ldots ,r;{\rm \; }i=1,2,\ldots ,n$) in the equality has $(n+1)$ independent variables: $T$, $P$, and the number of concentrations of chemical species included in each phase, which is ${n}-1$. (There are $n$ components included in the system. However, it is only $({n}-1)$ ratios that are independent.) 

   The equality (\ref{eq4}) consists of ${n}({r}-1)$ simultaneous equations, including $\{2+{r}({n}-1)\}$ unknown quantities. Thus, for the equations to have a solution, the number of equations should necessarily be less than the number of unknown variables. That leads to $n(r-1)\leqslant 2+r(n-1)$. Therefore, one finds the following inequality concerning the phase rule~\cite{1},
\[r\leqslant n+2.\] 
The theorem can be rephrased in the following way. In an equilibrium state in nature, more than $({n}+2)$ phases cannot exist at the same time. This is the Gibbs phase rule. If one applies the phase rule to a single component system such as pure water, then $n=1$, that leads $r\leqslant 3$. That proves non-existence of the ``second critical point'' in the real world. 

   Some SCP supporters have argued saying ``well, we are talking about the \textit{metastable} state of water, not the equilibrium state of water. The Gibbs phase rule may not be applied to a metastable state of water.'' The argument somewhat makes sense, because a metastable state is a kinetic state, or a nonequilibrium liquid-state between the two equilibrium states, liquid and solid. Therefore, the state should eventually be relaxed to an equilibrium or ice, after a long time, or quickly if one applies a small perturbation. Of course, the Gibbs phase rule says nothing about the metastable state of liquid, but it just says there are only two fluid states, gas and liquid, in the equilibrium state of a single component system. The author believes that it is enough to prove the non-existence of the SCP, or the liquid-liquid phase transition, in a single component system. There is \textit{only one} spinodal line, thereby there is only one critical point, in a single component system.

  Other SCP-supporters have argued that they are characterizing the kinetic state of the metastable state of the liquid. The argument does make sense if it is really the case. There is a beautiful work to characterize the kinetic state of the metastable water by Angle's group, referred to as the ``strong-fragile'' transition~\cite{14}. The characterization is made in terms of the kinetic properties of liquid, such as the viscosity and the diffusion constant. The behavior of the Arrhenius plot of the kinetic properties was employed to characterize the two kinetic schemes, ``fragile'' and ``strong''. Unfortunately, the SCP by Poole et al. was not characterized in terms of the kinetic properties, but with the equilibrium properties such as pressure and volume. Thus, this has no relation to a transition of the kinetic states. It is this double exposure of the two concepts to characterize fluid states, the critical point in the equilibrium state and the strong-fragile transition in the kinetic state, that makes the SCP conjecture equivocal.
  
  Some readers of the present paper may have a question concerning the multiple phases appearing in solid states of matter, for examples, the multiple crystal structures of ice. How does the Gibbs phase rule reconcile with such phenomena? The question can be answered in terms of the order of phase transition. In the case of the phase transition to which the Gibbs phase rule is applied, the first derivative of the Gibbs free energy, such as temperature, density, pressure, becomes discontinuous. In that respect, the phase transition is referred to as the first-order transition. On the other hand, in the case of crystal structure, the first derivative is continuous, but the second derivative, such as heat capacity and compressibility, becomes discontinuous. Such a phase transition is called the second-order transition. In order to characterize the second order transition, another physical parameter, or the crystal symmetry, should be introduced \cite{1}.  In any case, the critical point in the phase diagram is concerned with the first derivative of the free energy, to which the Gibbs phase rule should be applied strictly.

\section{On the second critical point found by the molecular simulations}

It is the MD simulation that has claimed existence of the second critical point in pure liquid water~\cite{3,5,6,7,8}.  Thus, there may be a naive question why the molecular simulation could find the second critical point in the pure liquid phase. Here, we try to give a possible answer to the question. The molecular dynamics simulations themselves cannot give a rigorous proof for robustness of their results. It is because they cannot meet the two requirements for the theoretical model to be meaningful as a \textit{thermodynamic} system. 

  Such first  requirement is the \textit{thermodynamic limit}~\cite{15,16}. Suppose, that a system having $N$ molecules are included in a container, the volume of which is $V$. Then, the thermodynamic limit is formulated as $\mathop{\lim }\limits_{N\to \infty ,V\to \infty } (N/V)=\rho $, where $\rho $ is constant, identified as average density of the system. The existence of the limit was proved mathematically based on the statistical mechanics for quite general class of systems. The requirement says that any sensible model of a natural system should have an infinite number of molecules involved in an infinitely large volume of container, keeping the density of molecules constant. Here, ``infinitely large'' means that the system is sufficiently large so that any effect from the system size on the thermodynamic property can be neglected. There are two effects conceivable, depending on the system size; the effect from wall of container and the number of molecules included in the container. Of course, the two are mutually related to each other. 

   It is the effect of the wall that has been concerned by the community of molecular simulation of liquid in the earlier stage of its development, since the possible number of molecules to be simulated are so small like $\sim 10^3$ at most in that period, due to the limited computer power. It is obvious that the effect from the wall may not be neglected, since the portion of molecules in contact with the wall is not so small to be neglected. Thus, this is not a uniform liquid system any more. In order to reconcile with this problem, the periodic boundary condition was invented in the community.

   The periodic boundary condition does not solve the other problem of the molecular simulation, concerning the system size including the number of molecules and the size or volume of container. In fact, this is the real problem to be solved in order for a theory or for a model to be able to attack the question related to the behavior of thermodynamic variables in the phase diagram. In order to make the discussion simpler, let us use the theory developed by Ornstein and Zernike, which is employed by many authors to discuss the behavior of fluid phases in the phase diagram, the phase equilibrium and the critical phenomena~\cite{17,18,19}.

   The critical point in the phase diagram is characterized by the point at which the derivative of the pressure with respect to volume is zero, or 

\begin{equation} 
	 \left(\frac{\partial P}{\partial V} \right)_{T} =0.
 \end{equation} 
It is also identified as the point where the isothermal compressibility diverges, that is

\begin{equation}  
	\chi _{T} \equiv -\frac{1}{V} \left(\frac{\partial V}{\partial P} \right)_{T} \to \infty.  
\end{equation} 
On the other hand, the compressibility is phenomenologically related  to the density fluctuation of fluid by

\begin{equation} 
	 \chi _{T} \propto \left\langle \rho -\left\langle \rho \right\rangle \right\rangle ^{2} . 
 \end{equation} 
Therefore, the divergence of  compressibility at a critical point is synonymous to say that the density fluctuation is taking place in a macroscopic scale. 

   In order to examine whether a molecular dynamics MD simulation can describe the critical point or not, one should have a microscopic expression of the compressibility that relates the thermodynamic quantity to the microscopic property such as intermolecular interactions. Such an expression is derived for both simple liquids and molecular liquids such as water, 

\begin{equation} 
	\chi _{T} \sim \int _{0}^{L}h(r)4\piup r^{2}  \rd r ,
	\label{eq8}
\end{equation} 
where $L$ is a parameter that measures the size of the container in which the molecules are contained. In the thermodynamic limit, $L=\infty $, obviously. The function $h(r)$ is a microscopic version of the density fluctuation, expressed by

\begin{equation} 
	 h(r)=\frac{1}{\left\langle \rho \right\rangle ^{2} } \left\langle \left(\rho (r)-\left\langle \rho \right\rangle \right)\left(\rho (0)-\left\langle \rho \right\rangle \right)\right\rangle,  
\end{equation} 
where $\rho(0)$ and $\rho(r)$ denote the density of molecules at the origin and at the radial distance ${r}$ from the origin in the spherical coordinate space, and $\left\langle \rho \right\rangle $ is the average density.

   The asymptotic behavior of $h(r)$ at $r\to \infty $ was investigated by G. Stell, and by M. Ohba and K.~Arakawa for the polar liquids described by the RISM theory, which leads to:

\begin{equation}  
	h(r)\sim \frac{1}{r} \exp \left(-r/\xi \right){\rm \; \; \; at\; \; }r\to \infty,
	\label{eq10}  
\end{equation} 
where $\xi$ is called ``correlation length'' which is a measure of persistent length of the density fluctuation~\cite{18,19}. Thus, with equation (\ref{eq8}), the compressibility will become asymptotically,
\begin{equation}
\chi _{T} \sim \int _{0}^{\infty }r\exp \left(-r/\xi \right) \rd r,                    
 \label{eq11}
 \end{equation}
which surely diverges upon $\xi \to \infty $.

   If one applies the formula to calculate the compressibility from a MD simulation, the equation should be modified to be something like,
\begin{equation}
\chi _{T}^{\text{MD}} \sim \int _{0}^{L}r\exp \left(-r/L\right) \rd r,  
\label{eq12}                  
\end{equation}
because the upper bound of integral should be around the box size, and the correlation length cannot exceed the box size. The integration can be readily carried out, and gives

\begin{equation}  \chi _{T}^{\text{MD}} \propto L^{2}.  \end{equation} 
The result indicates that the compressibility from MD does not diverge at any point of the phase diagram, neither the first critical point nor the ``second critical point''.

   Thus, in order to be able to find the critical point by MD, the size of the system should be large enough so that the correlation length diverges, that is a macroscopic scale like a glass of water. It is obvious that the molecular simulation, by which the \textit{second critical point} was claimed to be observed, is far from the thermodynamic limit. Therefore, it is a big mystery how the scientists could find the second critical point in their MD simulation for the system that contains as little as $\sim 10^3$ water molecules, which is desperately far from the thermodynamic limit. A possible answer to the question may be provided in terms of a \textit{structural change} or an ``isomerization'' of a nanoscale molecular-cluster from low density to high density. The essential point of the argument will not change if one increases the system size to $\sim 10^{8}$  molecules which is about the biggest size so far among the simulation community, since it is still a tiny cluster of molecules if one compares it with a real system which has $\sim 10^{23}$ molecules. 

   There is another problem to be concerned with when one analyzes the metastable phase in the supercooled region, that touches the Ergordic theorem in the statistical mechanics~\cite{20}. A MD simulation is a method to trace the trajectory of a many-body system in the phase space, and any thermodynamic quantity, such as pressure, temperature, density, and so on, is an average of the corresponding mechanical quantities over the trajectory: for example, temperature is essentially an average of the kinetic energy, and pressure is an average of a quantity called ``virial'' over a trajectory~\cite{17,21}. The Ergordic theorem imposes the following requirement to the average over the trajectory,

\begin{equation}  
	\left\langle A\right\rangle =\mathop{\lim }\limits_{\tau \to \infty } \frac{1}{\tau } \int _{0}^{\tau }A(q,p,t )\rd t ,
	\label{14}
\end{equation} 
where $A(q,p,t)$ is a mechanical quantity fluctuating around its most probable point in the phase space, with the multiple decay time depending on its mode: vibrational mode, rotational mode, collective mode, and so on~\cite{21}. The mode that concerns the macroscopic fluctuation around the critical point in the phase diagram will have a phenomenological decay time, like seconds or hours. Now, let us think of water in a supercooled region of the phase diagram which is in a state of low temperature. The velocity of molecules in the system will take a Maxwell distribution around an average velocity that is very slow. Therefore, it will make the sampling of the trajectory very time-consuming, even for a small system including only $\sim 10^3$ molecules. If one tries to make the system larger in order to meet the requirement of the thermodynamic limit, the method will break down quickly. If one replaces the MD simulation with the Monte Carlo (MC) simulation, the problem will not be solved, since a MC simulation is just another way of sampling the phase space of a system, but just confined to the configuration space (or positional space) assuming that the momentum is governed by an equilibrium distribution, or the Maxwell distribution. 

   Anyway, it can be readily imagined in the light of the Gibbs phase rule that the SCP or the ``no man's land'' departs from the real nature farther and farther as the molecular simulation approaches  the two limits, i.e., the thermodynamic and Ergordic limits.

\section{On the ``experimental evidences'' claiming the second critical point} \label{sec4}

Since the ``stability limit conjecture'' in fluid phase was proposed by Stanley and his coworkers,  many experimental ``evidences'' have appeared claiming the existence of the SCP~\cite{9,10,11}: the equilibrium of two liquid phases in the black phosphorus at high pressure and high temperature~\cite{9}; the phase separation between two liquid phases consisting of triphenyl phosphite~\cite{10}; the equilibrium between two ``liquid phases'' of water in protein measured by the neutron scattering~\cite{11}. There is no question in their experimental results concerning the existence of some sort of discontinuity in physicochemical properties and/or in their derivatives. However, it may be too hasty to relate such discontinuities to the SCP, or the ``no man's land'' claimed by Stanley and his coworkers. This section is devoted to clarifying the physics behind the phenomena that they refer to as SCP. In the course of clarification, a new concept of the liquid-liquid phase transition is proposed, which may be referred to as ``liquid-liquid phase-transition conjugated with an electronic-structure change of molecules''. For that purpose, it will be worthwhile to describe the chemical potential at the \textit{molecular} level, rather than at the \textit{phenomenological} level. 

   The chemical potential at the molecular level can be separated into two terms so that
\begin{equation} 
             \mu _{n}^{\Phi } =\mu _{n}^{\text{intra}} +\Delta \mu _{n}^{\Phi }, 
             \label{eq15}
\end{equation}              
where $n$ and $\Phi$ identify the chemical species and the thermodynamic phase, respectively, as in equation~\ref{eq4}. $\mu _{n}^{\text{intra}}$ is an intramolecular property that is essentially determined by the atomic composition and structure of a molecule, which  in its turn is determined by the electronic structure of the molecule. We refer to this part of the chemical potential as ``intramolecular chemical potential''. The second part of the chemical potential, $\Delta \mu _{n}^{\Phi }$, is referred to as the \textit{excess chemical potential}. The excess chemical potential depends essentially on the state of assembly, or the thermodynamic phase, which is casually called ``solvation free energy''. The thermodynamic equilibrium between two phases, I and II, is defined by the equality, $\mu _{n}^{I} =\mu _{n}^{II} $, for all species, denoted by $n$, included in the system.

The physical reason why some liquid mixtures show the phase separation is because the chemical potential of each component should be equal between the two phases by definition. The difference in the intramolecular part of the chemical potential should be compensated by the excess part of the chemical potential, which depends on the state of molecular assembly, or the thermodynamic phase, in order to meet the condition of the thermodynamic equilibrium. The equilibrium between the two fluid phases will be attained through the diffusion process driven by the gradient of the \textit{overall} chemical potential. In the conventional case of a single-component liquid such as water, it is considered that the intramolecular part of chemical potential $\mu _{n}^{\text{intra}}$ does not depend on the thermodynamic condition, or phase. Therefore, the phase behavior of pure liquids should be determined thoroughly by the excess part of the chemical potential, or $\Delta \mu _{n}^{\Phi }$, which of course does not induce any phase-transition or separation in consistency with the conventional Gibbs phase rule. This is the case of pure water as was clarified in the preceding sections.

Then, what is the physics behind the liquid-liquid phase transition or separation shown in some class of liquids consisting of a ``single'' chemical component, such as the phosphorous liquid and the liquid of triphenyl phosphite? The both cases can be explained in terms of the Gibbs phase rule, but with the concept of \textit{liquid-liquid phase-transition conjugated with an electronic-structure change of molecules}. The new concept assumes that both the intramolecular part $\mu _{n}^{\text{intra}} $ and the excess part $\Delta \mu _{n}^{\Phi } $of the chemcial potential depend on the thermodynamic state, density, temperature, and/or the pressure of the system, and those two interplay with each other. Actually, such a concept has been developed in the community of solution chemistry in order to explain the solvent effect on the electronic structure of a molecule in the past few decades.

Let us  briefly introduce one of the theories developed by us, which is referred to as RISM-SCF theory~\cite{22,23,24}. In the theory, the intramolecular part $\mu _{n}^{\text{intra}} $ of the chemical potential is defined as the electronic energy of the molecule, described by the so-called ``solvation Fock operator,'' which consists of two terms, i.e., the electronic Hamiltonian of the molecule and the interaction with the electrostatic \textit{reaction-field} exerted by environmental \textit{solvent}. The reaction field is evaluated using the site-site radial distribution function (SSRDF) obtained from the RISM theory, as the statistical weighting factor for taking the ensemble average. However, the RISM calculation requires the classical Hamiltonian, in which the electrostatic energy is described by the classical Coulomb-interaction between the partial charges which, of course, are the classical analogue of the electronic distribution. It is the basic idea of the theory to solve the Hartree-Fock type equation and the RISM equation iteratively until both the electronic distribution and SSRDF converge. The excess part of the chemical potential, $\Delta \mu _{n}^{\Phi } $, can be calculated from the converged result of the SSRDF. It is this process of iteration that reproduces the interplay of intramolecular and excess parts of the chemical potential.

   It is well known that the phosphorous liquid undergoes a liquid-liquid phase transition upon the rising temperature. The phenomenon is also highlighted by the supporter of SCP as an experimental evidence for the existence of the liquid-liquid phase transition in a pure liquid~\cite{9}. However, in this case, the liquid cannot be regarded as a single component system. As the analysis of the neutron diffraction data by the authors indicates, a sort of chemical reaction from tetramer of phosphorus atoms to larger aggregate or polymer is taking place upon the increasing pressure. The reaction should be associated with a dramatic change in the electronic structure, and the intramolecular part of chemical potential $\mu _{n}^{\text{intra}} $ of the phosphorus atoms in the two states of aggregation, a tetramer and a polymer, should not be the same anymore. This point was proved theoretically by Morishita based on the first principle MD simulation~\cite{25}. The reason why they  used the first principle MD is because the electronic structure of phosphorus atoms changes depending on pressure. Thus, we should regard the system as a mixture of two different species with different intramolecular chemical potentials, $\mu _{n}^{\text{intra}}$, the ratio of which depends on the pressure. If this is the case, there is no mystery in the phenomenon that shows the transition between the two \textit{liquid phases} in the light of the Gibbs phase rule. The physics can be explained by the concept of \textit{liquid-liquid phase-transition conjugated with an electronic-structure change of molecules.}

   The second example of experimental observations is the liquid-liquid phase transition in liquid triphenyl phosphite, reported by Tanaka et al.~\cite{10}. They  carried out complex (AC) heat capacity measurements with a fast scanning differential calorimetry on the sample, and found the \textit{liquid-liquid phase transition}. Based on the results, the authors  claimed that the result is another experimental evidence for the liquid-liquid phase transition in a single component liquid, originally found by Pool et al. by means of the molecular simulation for pure water. However, now, we should ask a serious question regarding the interpretation of their results in the light of the Gibbs phase rule: ``Is the liquid triphenyl-phosphite really a single component system, identified with a single intramolecular chemical potential?'', ``Doesn't it change the structure of molecules along the course of the measurement changing pressure and temperature?''. If it does, the liquid cannot be regarded as a single component system. Rather, it should be considered as a mixture of liquids, each component of which is identified by its own intramolecular chemical potential $\mu _{n}^{\text{intra}} $. Apparently, the chemical compound treated by the experiment looks quite flexible having a lot of rotational freedom of the three phenyl-rings around the O-C bonds. Further, the density of those ``isomers'' seems to be quite different from each other due to the packing of the phenyl ring. Therefore, it is quite possible that the molecules undergo \textit{isomerization} along the course of the change in pressure and temperature, which involves a dramatic change in the electronic structure of a single molecule. Thus, the physics of this phenomenon can also be explained by the concept of \textit{liquid-liquid phase-transition conjugated with an electronic-structure change of molecules.}

   Those experimental findings are of their own scientific interests themselves, because they are concerned with a novel class of liquid-liquid phase transition in which the phase transition and chemical reactions interplay with each other. 

    The case of water confined inside protein, studied with the neutron scattering, seems to be even less relevant to the liquid-liquid phase transition~\cite{11}. The experiment itself is quite popular among the biophysicists featuring the elastic coherent neutron scattering measurement on powder samples of protein~\cite{27}. The logarithm of the structure factor plotted against the square of wave vector becomes linear with negative slope at the small wave vector region. From the gradient of the plot, one finds the mean square displacement (MSD) of protein, or $\sum _{\alpha }\left\langle \Delta R_{\alpha }^{2} \right\rangle  $, which is a measure of structural fluctuation of protein, and it changes linearly with temperature. The MSD plotted against temperature exhibits an abrupt change in its slope around $T=230$~K~\cite{27,28}. The change was interpreted in terms of the liquid-liquid phase-transition by Chen et al.~\cite{11}. However, two serious questions are raised to the interpretation of the experimental result. Is the water inside protein so abundant as to be called a ``liquid phase''? Does the experiment principally probe the density fluctuation of the solvent? The answer to the both questions should be definitely ``No''. Actually, the two questions touch the important issues in biophysics: the function, stability, and structural fluctuation of protein. It is true that water plays a crucial role for a protein to perform its function. However, water  plays such roles not as a ``phase'', but as a molecule, for example, as substrates and cofactors of  enzymatic reactions~\cite{29}. Water plays an essential role for stabilizing and/or destabilizing the protein structure, but again those water molecules recognized inside protein  play their roles as a \textit{molecule} not as a \textit{phase}~\cite{29}.

\section{Conclusion and perspective}

Non-existence of the SCP  and the liquid-liquid phase transition in the supercooled region of water was proved theoretically based on the Gibbs phase rule. It was argued based on the statistical mechanics of liquids that the ``second critical point'' identified by the molecular simulation is not a real critical point but an artifact created by the method due to its incapability to cover the thermodynamic limit. Any theoretical attempts by means of the equilibrium statistical mechanics to realize SCP will fail due to the following reasons: firstly, SCP does not exist in nature according to the Gibbs phase rule as proved in the present paper; secondly, the supercooled region in the phase diagram is a non-equilibrium phase, so that the (equilibrium) statistical mechanics cannot be applied. 

   It is suggested that the singularity in the supercooled water, observed by the simulation, may be related to the non-equilibrium process such as the \textit{strong-fragile transition,} proposed by Angel~\cite{14}. Thus, it is essential to describe the phenomenon in terms of non-equilibrium or kinetic properties such as the viscosity and diffusion constant. The problem may be solved by the molecular simulation with a sufficient care concerning the Ergoric theorem. The problem may be also challenged by the statistical mechanics of molecular liquid such as the RISM theory combined with the generalized Langevin equation~\cite{29}. 

   The existence of the liquid-liquid phase transition in some ``single component'' systems in confined space such as pore, capillary, and solid surfaces, claimed by some experimental studies, is apparently due to misinterpretation of the physics behind the physicochemical processes. The processes interpreted as ``liquid-liquid phase transition'' are the phenomena described in terms of the \textit{absorption} and/or \textit{adsorption} in the conventional surface science. Such phenomena should not be interpreted as the phase transition of single-component liquids, but as processes induced by the molecular interaction between liquid and surface. A statistical-mechanical treatment for liquids confined in porous media was proposed by Kovalenko and Hirata based on the replica-RISM theory~\cite{30}. It was found that the porous media gives a strong effect upon water to produce a phase diagram that is entirely different from the pure water. Water molecules confined in a cavity of protein seem to be misinterpreted also from the viewpoint of SCP. In the community of life science, such phenomena are the process referred to as ``molecular recognition'' that plays crucial roles for activities of protein, including the enzymatic reaction~\cite{31}. A theoretical characterization of the molecular recognition was made by the author's group based on the RISM/3D-RISM theory~\cite{32}.  

   Liquid-liquid phase transitions exhibited by phosphorus and Triphenyl-Phosphite liquids are the other cases that claim the existence of the second critical point in liquids consisting of a single chemical component~\cite{9,10}. It was argued in the present paper that such liquid systems should not be regarded as single-component systems, but as a solution consisting of two chemical components identified with the different \textit{intra molecular} chemical potentials. The intramolecular part of the chemical potential of a molecule is determined by the structure of the molecule, which  in its turn is determined by the electronic structure. In such molecules as phosphorus and Triphenyl-Phosphite, the \textit{intra}molecular part of the chemical potential is significantly influenced by interactions with the surrounding molecules, or ``solvation'', which depends on the thermodynamic property of the system, especially on the density. It was hypothesized in the present paper that the phase transition is induced by the \textit{conjugation with an electronic-structure change of the molecules.}

The hypothesis may be proved experimentally by means of the molecular spectroscopy. If the electronic spectrum, for example, changes significantly along the course of the phase transition, it indicates that the molecular structure before and after the transition is not the same. Therefore, it will be an unambiguous evidence that the intra molecular part of the chemical potential is conjugated with the process of the phase transition. 

 A theoretical proof of the hypothesis requires a methodology that combines the statistical mechanics of molecular liquids with the electronic-structure theory, since the thermodynamic process of the phase transition interplays with the intramolecular quantum process. The RISM-SCF theory founded by the author and his coworkers is one of the candidates to solving the problem, although a further development of the methodology may be required~\cite{22,23,24}.

\ukrainianpart
\title[Чи дійсно існує у природі друга критична точка води?]%
{Чи дійсно існує у природі друга критична точка води?%
}

\author[Ф. Хірата]{Ф. Хірата}

\address{Інститут молекулярних досліджень, Оказакі, Аічі-444-8585, Японія
}
\makeukrtitle
	\begin{abstract}
		
		За останнє десятиліття літературний вираз ``нічийна земля'' заповнив собою наукові статті. Цей вираз використовується для опису метастабільної області у фазовій діаграмі, яку неможливо оцінити експериментально. На основі моделювання методом молекулярної динаміки (МД) стверджувалося, що на ``нічийній землі'' існує критична точка або друга критична точка (ДКТ), і це викликало велику суперечку серед науковців. Дана стаття доводить, що гіпотеза ДКТ повністю суперечить строгим термодинамічним співвідношенням, які відомі як \textit{правило фаз Гіббса}.
		Причина, чому МД моделювання виявило помилкову ДКТ полягає у тому, що цей метод порушує вимогу, якій повинні задовольняти всі співвідношення статистичної механіки для точного відтворення термодинаміки. І основним питанням тут є термодинамічна границя. З'ясовано, що таке фазовий перехід ``рідина-рідина'' і ДКТ у чистих рідинах, які \textit{виявлені} в результаті моделювання та деяких експериментів. Запропонована нова концепція для пояснення фізики фазового переходу ``рідина-рідина'', що спостерігається експериментально в \textit{однокомпонентних} рідинах.	
		
\keywords термодинаміка, правила фаз Гіббса, друга критична точка
		
	\end{abstract}
\end{document}